\documentclass[]{cupconf}
\input{psfig}

  \checkfont{eurm10}
  \iffontfound
    \IfFileExists{upmath.sty}
      {\typeout{^^JFound AMS Euler Roman fonts on the system,
                   using the 'upmath' package.^^J}%
       \usepackage{upmath}}
      {\typeout{^^JFound AMS Euler Roman fonts on the system, but you
                   dont seem to have the}%
       \typeout{'upmath' package installed. cupconf.cls can take advantage
                 of these fonts,^^Jif you use 'upmath' package.^^J}%
      }
  \else
  \fi

  \checkfont{msam10}
  \iffontfound
    \IfFileExists{amssymb.sty}
      {\typeout{^^JFound AMS Symbol fonts on the system, using the
                'amssymb' package.^^J}%
       \usepackage{amssymb}%

      }{}
  \fi

  \IfFileExists{amsbsy.sty}
    {\typeout{^^JFound the 'amsbsy' package on the system, using it.^^J}%
     \usepackage{amsbsy}}
    {}

\newsavebox{\astrutbox}
\sbox{\astrutbox}{\rule[-5pt]{0pt}{20pt}}

\title[Massive Star in the Local Group]
{Massive Stars in the Local Group: Star Formation and Stellar Evolution}

\author[Philip Massey]%
{Philip Massey$^1$}
\affiliation{$^1${Lowell Observatory, 1400 W. Mars Hill Rd., Flagstaff, 
	AZ\ \ \ 86001, USA}}

\pubyear{2003}

\begin{document}

\maketitle

\begin{abstract}

The galaxies of the Local Group that are currently forming stars can serve
as our laboratories for understanding star formation and the evolution
of massive stars.  In this talk I will summarize what I think we've learned
about these topics over the past few decades of research, and briefly mention
what I think needs to happen next.

\end{abstract}

\firstsection 
\section{Introduction}
\label{Sec:intro}

My talk today will be restricted to giving a brief introduction to the study
of massive stars in the Local Group; I'll begin by discussing why I think
the subject is important, and giving you a few of the complications and
caveats.  I'll spend most of my time then talking about what I think we've
learned, first about star formation (stories of star formation, the initial
mass function, and the upper mass cut-off) and second about the evolution of
massive stars (including Luminous Blue Variables, Wolf-Rayet stars, and red
supergaints).  Finally I'll conclude with a brief discussion of where I
think we need to do next.  This talk is based in large part on an {\it Annual
Reviews of Astronomy \& Astrophysics} paper that I have coming out in October
(Massey 2003), and the reader is referred there for a more in depth analysis.
I have used this opportunity to update some of the figures and thoughts from
that, so hopefully the two will be somewhat complementary.  

Massive stars are extremely rare.  If we integrate a Salpeter (1955) initial
mass function (IMF) and allow for the relative lifetimes involved, we expect
that there are about a hundred thousand solar-type stars (i.e., stars of 1$(\pm 0.1) \cal  M_\odot$ for every 20$ (\pm 0.2) \cal M_\odot$ in the Galaxy.  There should
be over a million solar-type 
stars for every 100$ (\pm 10) \cal M_\odot$ star.
Yet, these stars have a disproportionate effect upon their environment:
they provide most of the mechanical energy input into the interstellar
medium through stellar winds and supernova (Abbott 1982);  they provide
most of the UV ionizing radiation of galaxies, plus power the far-IR
luminosities of galaxies through the heating of dust; and they serve
as the primary source of CNO enrichment of the interstellar medium
(Maeder 1981).

Why study massive stars in the more distant galaxies of the Local Group,
when there are so examples closer to home?  The primary reason is that
the metallicity of the gas out of which these stars are born differs by
a factor of nearly 20 amongst the galaxies currently actively forming stars
(Massey 2003 and references therein).  And metallicity should matter!
First,  we expect that the initial mass function will depend upon metallicity, as the Jeans length depends upon temperature, and the
temperature depends upon the metallicity of the star-forming cloud.
Understanding how star-formation processes (and the IMF) varies with
metallicity is important for interpreting the 
integrated properties of galaxies at large look-back times.  
Secondly, the stellar winds of massive stars are driven by radiation 
pressure in highly ionized metal lines, and the evolution of massive stars is dominated by the effects
of radiatively-driven stellar winds.  Thus we expect to see large differences
in the evolved massive star populations depending upon the metallicity of
the host galaxy.

And this is a wonderful time for such studies.  Increasing sophisticated
stellar evolutionary models are available through the kindness of both the
Geneva and Padova teams.  At the same time we have excellent observational
capabilities, such as the high resolution optical imaging and UV spectroscopic
facilities of {\it HST}.  There are large-format CCD cameras available on
4-m class telescopes around the world, and my own ``Local Group Survey" team
is busy producing {\it UBVRI} photometry of 300 million stars in 9 nearby
galaxies.  At the same time there are high through-put spectrographs on
large telescopes, such as the Blue Channel on the MMT, GMOS on Gemini, 
and DEIMOS on Keck.

\subsection{Some Difficulties and Caveats}
\label{Sec:SDC}

\begin{figure}
\psfig{figure=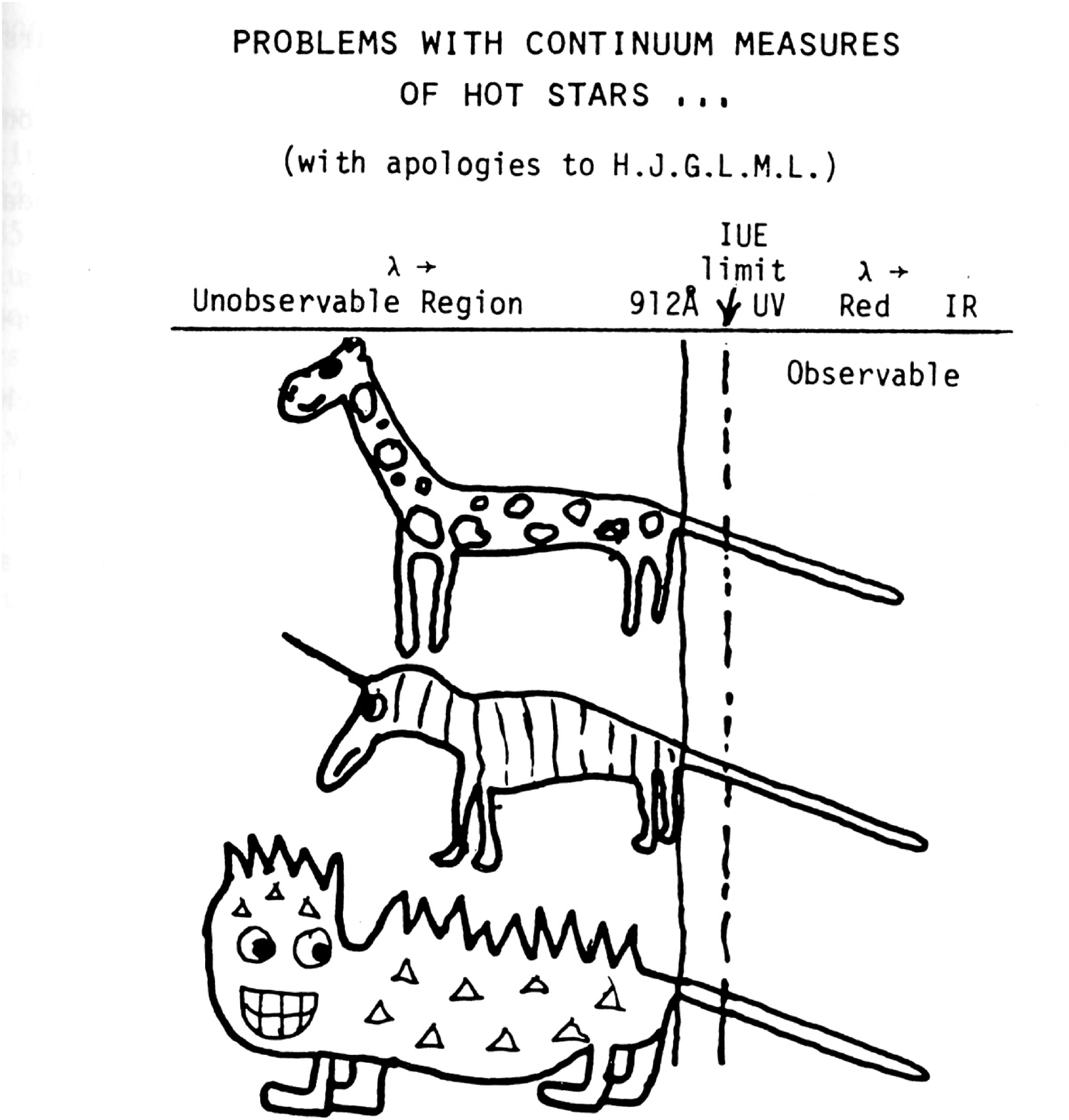,width=5.0truein}
\caption{For stars with high effective temperatures, the
{\it tails} of the Rayleigh-Jeans distribution gives little information
about the effective temperatures, and hence the bolometric luminosities or
masses of hot stars.  The figure is from Conti (1986) and is used with
permission. \label{fig:psc}
}
\end{figure}

Because massive stars have very high effective temperatures (30,000-50,000 $^\circ$K) while on the main-sequence, there are some unique problems in
characterizing the massive star population of a nearby galaxy.  Only a small
fraction ($<10$\%) of their light leaks out into the visible range, and because
what we see is so far down on the {\it tail} of the Rayleigh-Jeans distribution
that we may not recognize the actual beast (Figure~\ref{fig:psc}).  The
bolometric luminosity depends critically on the effective temperature, and
the deduced mass depends critically on the bolometric luminosity.  Massey (1998a)
gives explicit equations,  namely:
$$L\propto m^{2.0}$$
$$\Delta \log m = -0.2 \times \Delta M_{\rm bol}$$
$$\Delta {\rm BC} =-6.84 \times \Delta \log T_{\rm eff}.$$
A reddening-free index, such as $Q$, will have a dependence $\Delta{\rm BC}/
\Delta Q=33$ (note that the ratio is mistakenly inverted in Massey 1998a); i.e., a 0.10~mag uncertainty in $Q$ will lead to an
uncertainty in the BC (and $M_{\rm bol}$ 
of 3.3~mag and hence to an uncertainty of 0.7~dex in $\log m$.  This is not
going to lead to finding a meaningful IMF slope. 

Another implication of the high effective temperatures, and corresponding
large bolometric corrections, is that {\it the visually brightest stars are
not the most massive}.  A young 85$\cal M_\odot$ star O-type star will
be about 3~mags (a factor of 15) fainter than a 
25$\cal M_\odot$ A-type supergiant, although
the later is about 1.5~mag (factor of 4)
less luminous bolometrically.  The reader is referred to Figure 1 in 
Massey et al.\ (1995) and the corresponding discussion.

In both of these cases spectroscopy provides the answer to such ambiguities,
allowing accurate temperatures (and hence BCs, luminosities, and thus masses)
to be determined, and allowing the construction of meaningful H-R diagrams.

\subsection{Tests}
\label{Sec:Tests}

If indeed the Local Group is our ``astrophysical laboratory", then we must
be good scientists and design our experiments with some care.  Generally there
are two types of tests possible.

\begin{enumerate}
\item {\it Mixed-age tests,} where we compare the relative number of
one thing to another, such as the relative number of red supergiants (RSGs) to
Wolf-Rayet stars (WRs).  
The implicit assumption in such tests---not often stated,
and possibly not always realized---is that we are averaging over a sample
population to be have a completely heterogeneous mix of ages.  If this
assumption is invalid---if we are instead looking at a population with
a strong bias of 10~Myr (typical for a RSG) but for which 4~Myr old stars
(typical for WRs) are under-represented, then we are going to have some
problems in interpreting the results.

\item {\it Coeval tests,} where we study the main-sequence stars in a cluster
in order to derive a stellar IMF or determine the turn-off mass in a cluster
containing evolved massive stars.  We must test if indeed all the stars were
``born on a particular Tuesday" (Hillenbrand etal.\ 1993) or if instead
the age spread is significant.

\end{enumerate}

\section{What Have We Learned?}
\label{Sec:WHWL}

\subsection{Star Formation, the IMF, and the Upper-Mass Limit}
\label{Sec:IMFUML}

\subsubsection{Star Formation}
\label{Sec:SF}

Detailed spectroscopic and photometric studies of young clusters reveal very
different histories of star formation.  Herbig (1962) first suggested that
low- and intermediate-mass stars might form over a prolonged time in a cluster,
followed by the formation of high mass stars, which halts all star formation.
But not all clusters follow this paradigm.

Back before {\it HST} made M16 famous, Hillenbrand et al.~(1993) studied
the stellar content of the cluster and found that there were intermediate-mass
stars pre main-sequence stars with ages as young as a few times $10^5$ years.
Yet M16's massive star population has an age of two million years.  As
Hillenbrand et al.~(1993) conclude, ``...Thus the formation of O stars neither
ushered in nor concluded the star-formation process in this young complex.

By contrast, the prototype ``super star cluster" R136 in the 30 Doradus
region of the LMC does fit the standard model.
{\it HST} photometry and spectroscopy reveals that the
intermediate-mass population began forming 6~Myr ago, and stopped about 2~Myr years ago.  This cluster contains a very large number of extremely
massive stars that formed 1-2~Myr ago (Massey \& Hunter 1998).

These two clusters provide an interesting contrast in their stories of
star formation.  In very rich, dense clusters, such as R136, we find
that the formation of intermediate-mass stars is stopped shortly after the 
{\it formation} of the high-mass stars, due presumably to the effects
that their stellar winds have on the surrounding gas.  In a less rich
cluster such as M16, production of intermediate-mass stars is not halted
by the formation of the high-mass stars.  One might speculate that there
the intermediate-mass stars will continue to form until the first Wolf-Rayet
stars (with their strong stellar winds) are produced, or possibly will
continue until the first SNe.  Nevertheless, in both M16 and R136 the massive stars themselves were formed over a very short period of time ($<1$~Myr),
a fact that can prove very useful, as we'll see in Section~\ref{Sec:coeval}.

\subsubsection{The IMF of Massive Stars}
\label{Sec:IMF}

\begin{figure}
\psfig{figure=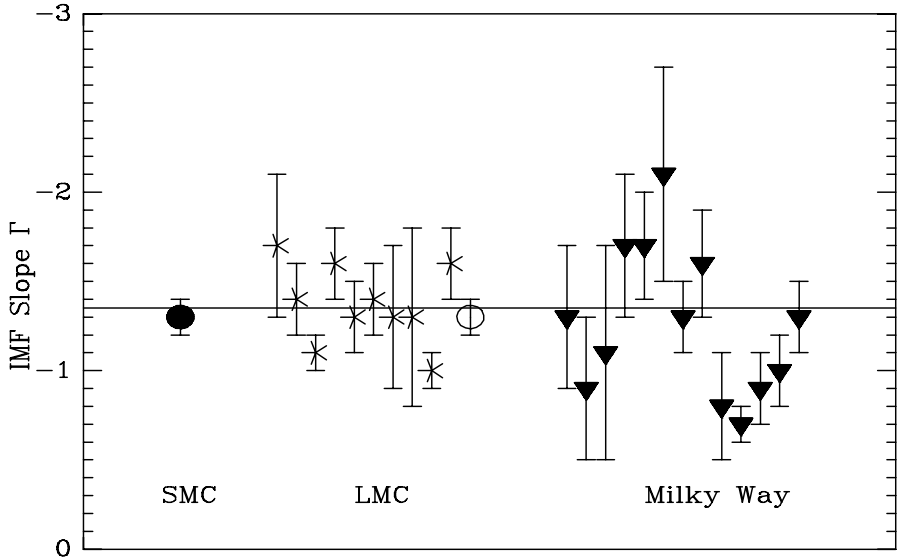,width=5.0truein}
\caption{The slope ($\Gamma$) of the 
initial mass function of OB associations 
and clusters in the SMC, LMC, and Milky Way is constant over a range of
4 in metallicity and 200 in stellar density.  The circle shows the IMF
of the R136 cluster in the LMC.  The Salpeter (1955) $\Gamma=-1.35$
value
is denoted by the solid line.
Based upon data in Massey (2003) and 
references therein. \label{fig:IMF}
}
\end{figure}

Studies of the stellar IMF of OB associations and clusters in the Milky Way
and Magellanic Clouds show no evidence for any effect with metallicity. The variations that are seen are observational and/or statistical in nature 
(Massey 1998a, Kroupa 2001).  In Figure~\ref{fig:IMF} we see that the 
{\it slope} if the IMF, $\Gamma$, is approximately Salpeter ($\Gamma=-1.35$)
over a factor of 4 in metallicity and 200 in stellar density.

This is actually somewhat surprising given that metals provide the
primary cooling mechanism in molecular clouds, and hence cloud temperatures (and thus the Jeans mass) should depend upon metallicity.  
But so far we have probed only a 
small range in metallicity: from the SMC to the Milky Way, the metal content
changes by only a factor of 3.7.  We could extend this to cover a range
of 17 if we were to push such studies further out in the Local Group
(WLM to Andromeda).  Our Local Group survey project is producing {\it UBVRI}
photometry of 300 million stars, but will require spectroscopy if we are
to determine IMFs and star formation histories.

\subsubsection{The Upper Mass Limit}
\label{Sec:UML}

One of the other important lessons that spectroscopy of the R136 stars
taught us (Massey \& Hunter 1998) is concerns the ``upper mass limit" seen
in clusters.  The highest mass stars in R136 were an unprecedented 150$\cal M_\odot$.  However, this is just what we would have expected from extrapolating
the IMF, as this is just where the IMF would peter down to a single star.
This suggests that the ``upper mass limits" we've so far encountered are
statistical, and not physical.  Whatever it is that limits the
ultimate mass of a star, we have yet to encounter it in nature.

\subsection{Evolved Massive Stars}
\label{EMS}

The evolved (He-burning) massive stars include:
\begin{itemize}
\item Luminous Blue Variables (LBVs): very luminous stars that undergo
occasional ``eruptions" (S Doradus, $\eta$
Carinae, the Hubble-Sandage variables).

\item Wolf-Rayet stars (WRs): stars with broad emission lines whose surface compositions are consistent with the equilibrium products of H-burning (WN types) or He-burning (WC types).

\item Red supergiants (RSGs).

\end{itemize}

\subsubsection{The Conti Scenario}
\label{Sec:Conti}

Conti (1976) first proposed that stellar winds (mass-loss) would result in
some of the chemical peculiarities observed for Wolf-Rayet stars.  The most
luminous stars (at a given metallicity) will have the highest mass-loss rates
and suffer the greatest fraction of mass loss.  A modern version of the
``Conti scenario" might look something like this:

\begin{centering}

$m>85\cal M_\odot$: O$\longrightarrow$LBV$\longrightarrow$WN$\longrightarrow$WC$
\longrightarrow$SN

$40>m>85\cal M_\odot$:O$\longrightarrow$WN$\longrightarrow$WC$\longrightarrow$SN

$25>m>40\cal M_\odot$: O$\longrightarrow$RSG$\longrightarrow$WN$\longrightarrow$WC$\longrightarrow$SN

$20>m>25\cal M_\odot$: O$\longrightarrow$RSG$\longrightarrow$WN$\longrightarrow$SN

$10>m>20\cal M_\odot$:
OB$\longrightarrow$RSG$\longrightarrow$BSG$\longrightarrow$SN

\end{centering}

Thus, an 85$\cal M_\odot$ star starts out as an O-type star, and then
becomes a luminous blue variable.  Extreme mass-loss during the LBV stage
then helps lead first to a WN-type Wolf-Rayet star (with the H-burning
products revealed at the surface), while subsequent evolution (and mass-loss)
results in a WC-type WR (with the more advanced He-burning products revealed
at the surface.  Stars of lower mass (20-25$\cal M_\odot$) might not have
enough mass loss for a star to evolve beyond the WN stage, while stars of
even lower mass might never go through a WR phase at all, but rather spend
their He-burning lives first as red supergiant (RSG) and then as a 
``second generation" blue supergiant (BSG), similar to the precursor of 
SN1987A.

Of course, we don't know how right this overall ``cartoon" version of massive
star evolution is, and in particular we don't know what the corresponding
mass ranges are for the various evolutionary paths.  We especially don't know
how these ranges vary with metallicity!

\subsubsection{LBVs}
\label{Sec:LBVS}

Hubble \& Sandage (1953) called attention to five irregular
variables in M31 and M33 that were (at times) among the brightest resolved
stellar objects in these galaxies. Photographic plates from Mt.~Wilson
extended back to 1916, which revealed episodic visual brightenings of several
magnitudes.  These objects were extremely luminous and blue, and had
spectra that were of intermediate F-type during maxima.  A footnote in
their paper suggested that the LMC star S Doradus might be similar.  The
connection to the Galactic stars $\eta$~Car and P~Cyg came later.
Conti (1984) coined the term ``luminous blue variables" to describe these
objects.

LBVs are extremely luminous bolometrically, and are found near the edge of
the observed upper luminosity limit in the H-R diagram.  They undergo
episodic bouts of high mass loss, during which they brighten visually,
while remaining roughly constant in bolometric luminosity.

In a series of papers in the late 1970s and early 1980s, Roberta Humphreys
demonstrated that there was an observed limit to the luminosities of stars
(see, for example, Humphreys \& Davidson 1979).  This limit decreases with decreasing
effective temperatures until $T_{\rm eff}=$10,000$^\circ$K, after which
it is nearly constant at $/log L/L_\odot \sim 5.7$ ($M_{\rm bol}=-9.5$,
corresponding roughly to 50$\cal M_\odot$.  It was understood that this
upper luminosity limit was {\it somehow} related to LBVs, since these
star occupied a narrow band near this limit.

It was very tempting to try to interpret the observed upper luminosity limit 
in terms of some fundamental physics, such as the Eddington limit.  
In the classic Eddington limit,
$$\frac{L}{M} = \frac{4\pi Gc}{\kappa}$$
where $\kappa$ is the flux-mean opacity (0.347 for electron scattering if
$T_{\rm eff}$ is high).  In that case $$\frac{L/L_\odot}{M/M\odot}=3.8\times10^4$$
This is similar to, but considerably greater than, the {\it observed}
$\frac{L/L_\odot}{M/M_\odot}$ ratio of the Humphreys-Davidson limit:
$$\frac{10^{5.7}L_\odot}{50\cal M_\odot} = 1.0\times 10^4$$
This is tantalizingly close, but the {\it classical} Eddington limit is
thus factors of several times higher than what is observed, and doesn't show the
same $T_{\rm eff}$ dependence.

Lamers \& Fitzpatrick (1988) used model atmospheres to explore where
radiation pressure and gravity are balanced if the full effects of metal-line opacities are included, rather than just the effects of electron scattering.
This turned out to completely explain the Humphreys-Davidson limit.  The
actual limit is a trough, descending to lower luminosity with decreasing
effective temperatures until a temperature of 10,000$^\circ$K is reached,
after which the opacity decreases and hence the allowed luminosity
increases again.  However, since stars evolve from higher temperatures
to lower, the effect is to produce a decreasing envelope followed by
a line at constant luminosity luminosity for $T_{\rm eff}<$10,000$^\circ$K.

This suggests then that LBVs are a normal phase in the lives of the most
massive stars.  As such stars try to evolve to cooler
effective temperatures,
their radii expand, their surface gravities get lower, and most importantly
their  atmospheric opacities increase and radiation pressure overcomes gravity, leading to
vastly increased mass loss.  These stars are stopped dead in their (evolutionary) tracks!  Such a star will dump a lot of material, heat up, and 
then try again.

This may neatly wrap up the whole issue of LBVs, but possibly not.
Kenyon \& Gallagher (1985) suggested that some LBVs were the result of
evolution in close binary systems, an argument partially supported by the
relative isolation of some LBVs in M31.  This has been given some boast
by the controversial claim that $\eta$~Car itself is a binary.  

\begin{figure}
\psfig{figure=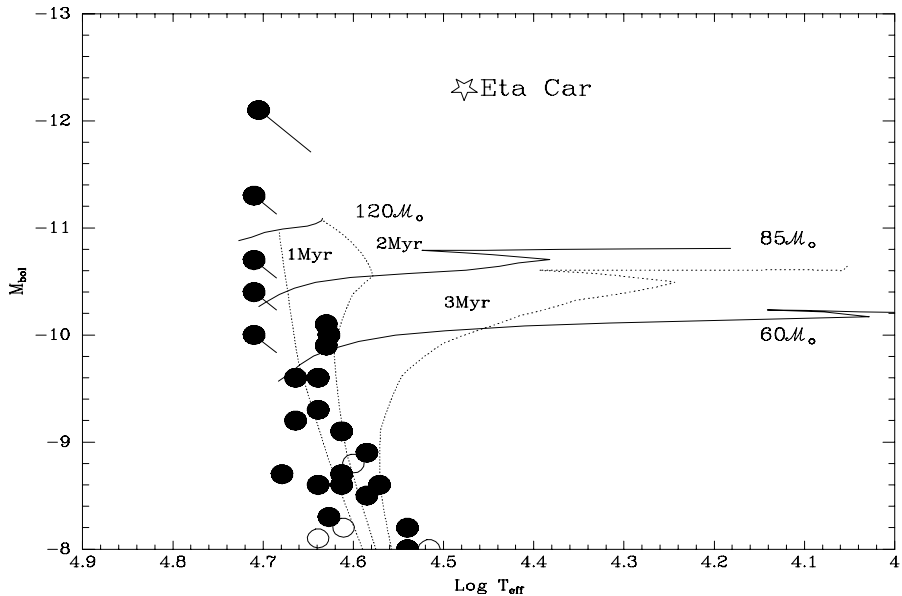,width=5.0truein}
\caption{The location of $\eta$ Car in the H-R diagram is just what we
expect if it happens to be a highly evolved massive star, as its bolometric
luminosity is slightly higher than the highest luminosity unevolved
massive star.  The highest mass stars are quite coeval, with an age
of about 1~Myr.  
This figure is based upon Figure 4 in Massey, DeGioia-Eastwood, \& Waterhouse (2001). \label{fig:eta}
}
\end{figure}

The problem with fully resolving the question of the origins of LBVs is
complicated by the fact that the bolometric luminoisities of LBVs are poorly
known, making their placement on the H-R diagram uncertain.  The real
exception to this is $\eta$~Car itself, where the surrounding gas and dust reprocesses the radiation to the IR, where it can be measured.  $\eta$ Car
is a member of the Tr14/16 complex, and when we place it on the H-R diagram
(Figure ~\ref{fig:eta})
we find something quite comforting: it is exactly where it should be, in the
sense of being just slightly more bolometrically luminous than the highest
luminosity unevolved star (Massey, DeGioia-Eastwood, \& Waterhouse 2001).
One is forced to conclude that even if $\eta$ Car is a binary, its binary
nature may be irrelevant to its LBV nature.

Another complication in understanding the origins of LBVs is that it is very
hard to gather meaningful statistics.  $\eta$ Car underwent its last major ourburst 
in 1830, and P Cyg underwent its last major outburst in 1655.  If these
stars were located in the Magellanic Clouds, would we even consider them LBVs?
One such Galactic example of an LBV-in-waiting  may be the star VI Cyg No.~12
(Massey \& Thompson 1991).  This star is one of the
visually most luminous stars known
in the Milky Way, it is surrounded by circumstellar material, but it has
not had a photometric ``episode" during the few decades its been known.
Given that Hubble \& Sandage (1953)
relied upon 40 years worth of observations just to identify the first 5
such stars in M31 and M33 (one of these 5 stars, Var~A, is no longer
even considered an LBV), it may require centuries to obtain firm statistics
on the number of LBVs in nearby galaxies.

\subsubsection{Wolf-Rayet Stars}
\label{Sec:WRS}

In 1867 two French astronomers, Wolf and Rayet, came across three
stars whose spectra showed broad, strong emission lines rather than the
absorption spectra that characterized other stars.  Beals (1930) correctly
identified the emitting atoms as ionized helium, nitrogen, and carbon.

WN-type Wolf-Rayets show lines of helium and nitrogen.
Abundance studies show that for WN stars we are seeing an atmospheric composition that reflects what the composition that the stellar core
would have during the main-sequence CNO hydrogen-burning cycle.
 (See discussion and
references in Massey 2003.)  WC stars show helium, carbon, and oxygen;
the latter two are the products of the triple-$\alpha$ helium-burning cycle.
Somehow, then, the outer layers of these stars have been stripped away,
revealing the nuclear burning products at the surface.  Prior to 1976, the
prevailing notion was that Wolf-Rayet stars were the product of evolution
in close binaries, with Roche-lobe overflow being responsible for this
stripping.  However, Conti (1976) noted that the ubiquitous presence of
stellar winds in hot stars provided an alternative explanation, and proposed
that O-type stars would naturally evolve first to WN-type and then to WC-type
Wolf-Rayets.

Since mass-loss depends upon metallicity, one expects that the relative 
number of WCs and WNs should vary from place to place.  Indeed, 
roughly equal number of WCs and WNs were known in the vicinity of the
sun, while in the lower metallicity Magellanic Clouds most of the WRs
are of WN type.  quantitative numbers provide the means for comparison with
evolutionary models.

One complication in finding unbiased samples of Wolf-Rayet stars is that
the WCs are much easier to find than are the WNs.  The strongest emission
line in WCs (CIII$\lambda 4650$) have a median line strength that is a
factor of 4 grater than that of the strongest emission line in WNs (HeII $\lambda 4686$).  My colleagues and I designed a set of 3 interference filters
in order to facilitate identification of WRs in nearby galaxies, and have used
this with some success in regions of nearby galaxies (see Massey \& Johnson 1998 for a recent summary of such studies).  Candidate stars are identified on
the basis of having a magnitude difference from an emission-band to
continuum-band filter, if that magnitude difference is significant compared
to the photometric errors.  Candidates are then observed spectroscopically.

\begin{figure}
\psfig{figure=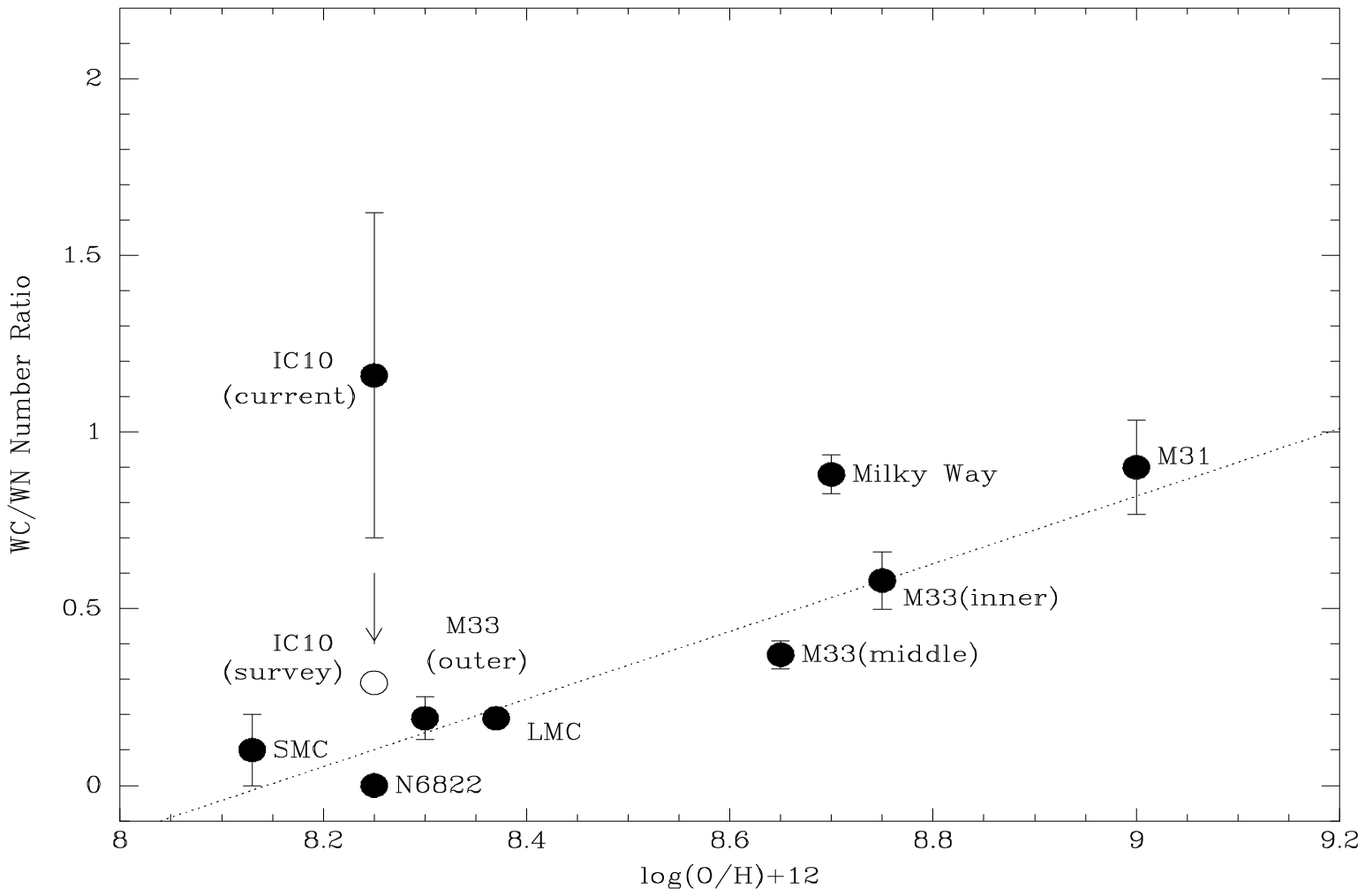,width=5.0truein}
\psfig{figure=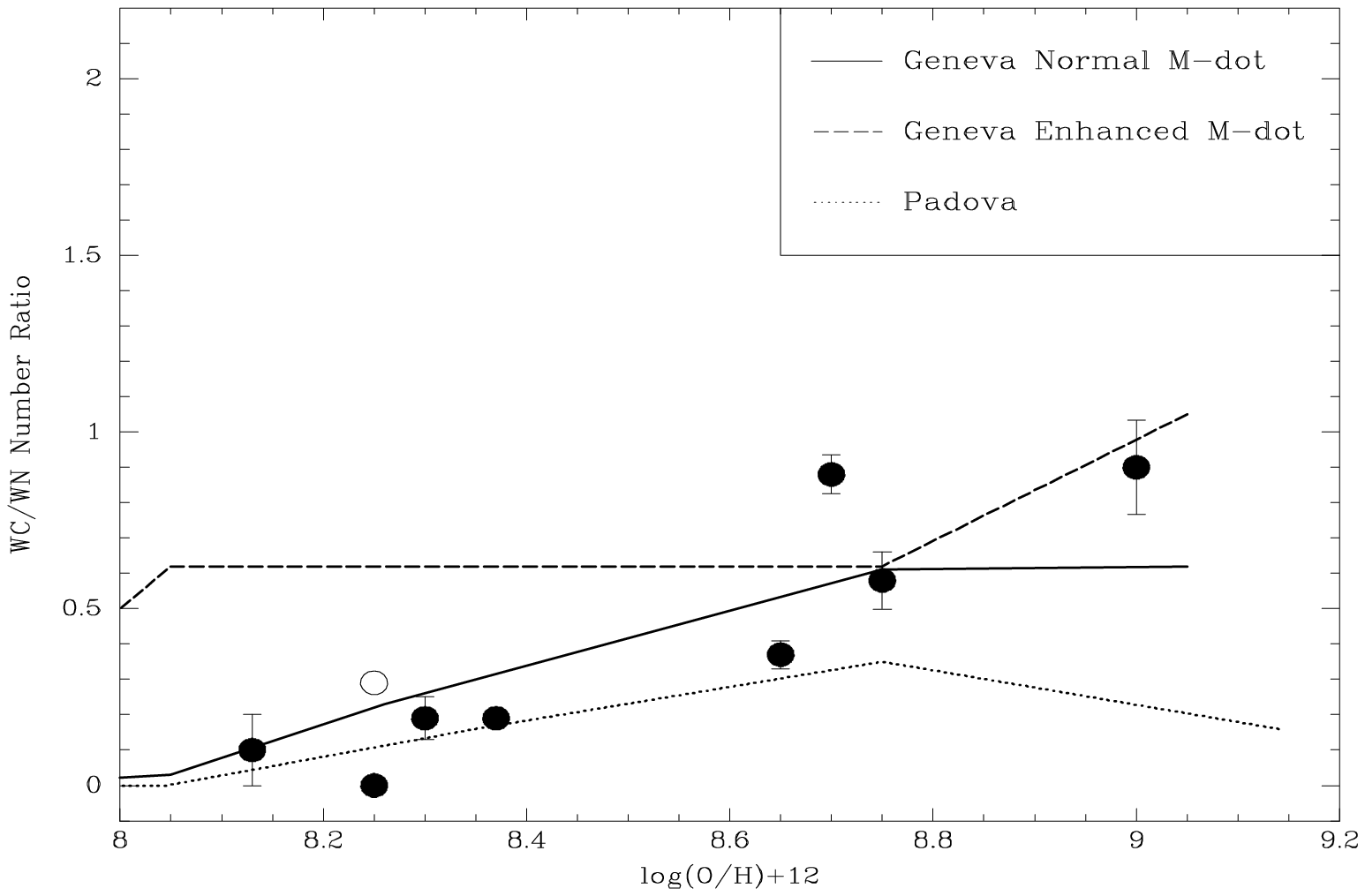,width=5.0truein}
\caption{The relative number of WCs and WNs show a strong trend with
metallicity among the Local Group galaxies (top).  The 
normal-mass loss models of the Geneva and Padova groups do a reasonable job at matching the data for all but the highest metallicities (bottom). The ``enhanced" mass-loss models of the Geneva group do not match the data at all. \label{fig:wrs}
}
\end{figure}

Such studies have revealed a strong trend in the relative number of WCs
and WNs with respect to metallicity. We show the data in Figure~\ref{fig:wrs}.
Two points deserve comments. First, the point for the Milky Way is higher
than the trend based upon the other galaxies, and one has to wonder if the
data for the region within 2~kpc is complete.  No systemic surveys for WRs
in the Milky Way have been carried out; instead, such stars were found
either as part of the HD survey (which would be very incomplete at a distance
of 2~kpc for WRs given typical reddenings, as noted by Massey \& Johnson 1998),
or by ``accidental" spectroscopic discovery as part of a study of stars in
a specific cluster, say.  IC10 poses a far more interesting discrepancy.
Massey \& Armandroff (1995) discovered 22 Wolf-Rayet candidates in this small
galaxy (by contrast, the SMC, which is twice the size, contains only 11 known
WRs), and argued that IC10 is the nearest example of a starburst galaxy.
One peculiarity though is that the number of WCs was about 2 times greater
than the number of WNs, while a ratio of 0.1 would be more in keeping with
its metallicity.  Various explanations have been advanced to explain this:
possible the IMF is top-heavy, or alternatively massive stars might have
formed within the same short time interval all across the galaxy and are
evolving in lock-step.  A recent deeper survey by Massey \& Holmes (2002)
suggests another explanation: they found a large number of additional
candidates, confirming two by spectroscopy.  If spectroscopic confirmation
of the remainder turns out the way they expect, then the number rati of
WCs to WNs may be normal in IC~10, but the galaxy may have a total WR content
of a 100 WRs, a surface density that would be about 20 times higher than that of the LMC.   In terms of its overall massive star population, IC10
must be more like an OB association, but on a kpc scale.

How well do the stellar models do in reproducing this trend?  We see in the
bottom panel of Figure~\ref{fig:wrs} that the normal-mass loss models
of the Geneva (Schaller et al.~1992; Schaerer et al.~1993; Charbonnel et al.~1993) and Padova (Fagotto et al.~ 1994; Bressan et al.~ 1993)
groups come relatively close to reproducing the
trends, although both predict too low a WC/WN at higher metallicities.
(One should note, though, that the statistics for M31 are not well 
established.)  The ``enhanced" mass-loss tracks of the Geneva group 
(Meynet et al.~1994) do
not match the data at all.  Massey (2003) has examined the historical basis
for the introduction of these ``enhanced" mass-loss models, and with the
advantage of 20-20 hindsight suggests that the problems these were trying to
solve were to some extent unreal.
Unfortunately, it is the enhanced mass-loss models that are commonly
used in starburst models (e.g., Schaerer \& Vacca 1998; Leitherer et al.\ 1999;
Smith, Norris, \& Crowther 2003).

\subsubsection{Red Supergiants}
\label{Sec:RSGs}

A long-standing problem in identifying the RSG content of nearby galaxies
has been the confusion by foreground Galactic dwarfs.  For instance,
Massey (1998b) found that roughly half of the sample of red stars seen
towards M33 by Humphreys \& Sandage (1980) were likely foreground Galactic
objects.  {\it BVR} two-color diagrams partially resolve this 
ambiguity, as RSGs will have redder {\it B-V} colors at a given
{\it V-R} color (Massey 1998b).  Spectroscopy of the Ca~II triplet lines
in the NIR is a powerful way of distinguishing Galactic disk dwarfs from
bona-fide extra-galactic RSGs given the heliocentric radial velocities of
most Local Group galaxies.

Massey (1998b) investigated the RSG content of three Local Group galaxies
(NGC 6822, M33, and M31), which span a range of 0.8~dex in metallicity.
He found that as the metallicity goes up, the fraction of high luminosity RSGs goes down, but that there is not a sharp cut-off in either $M_V$ or 
$M_{\rm bol}$.  This suggests that moderately high-mass stars (40$\cal M_\odot$)
become RSGs even at higher metallicities, but what changes is that the RSG
phase is shorter at higher metallicities, in accord with the suggestion of
Maeder, Lequeuex, \& Azzopardi (1980).  A recent study (Massey \& Olsen 2003)
has extended this work to that of the Magellanic Clouds.

\begin{figure}
\psfig{figure=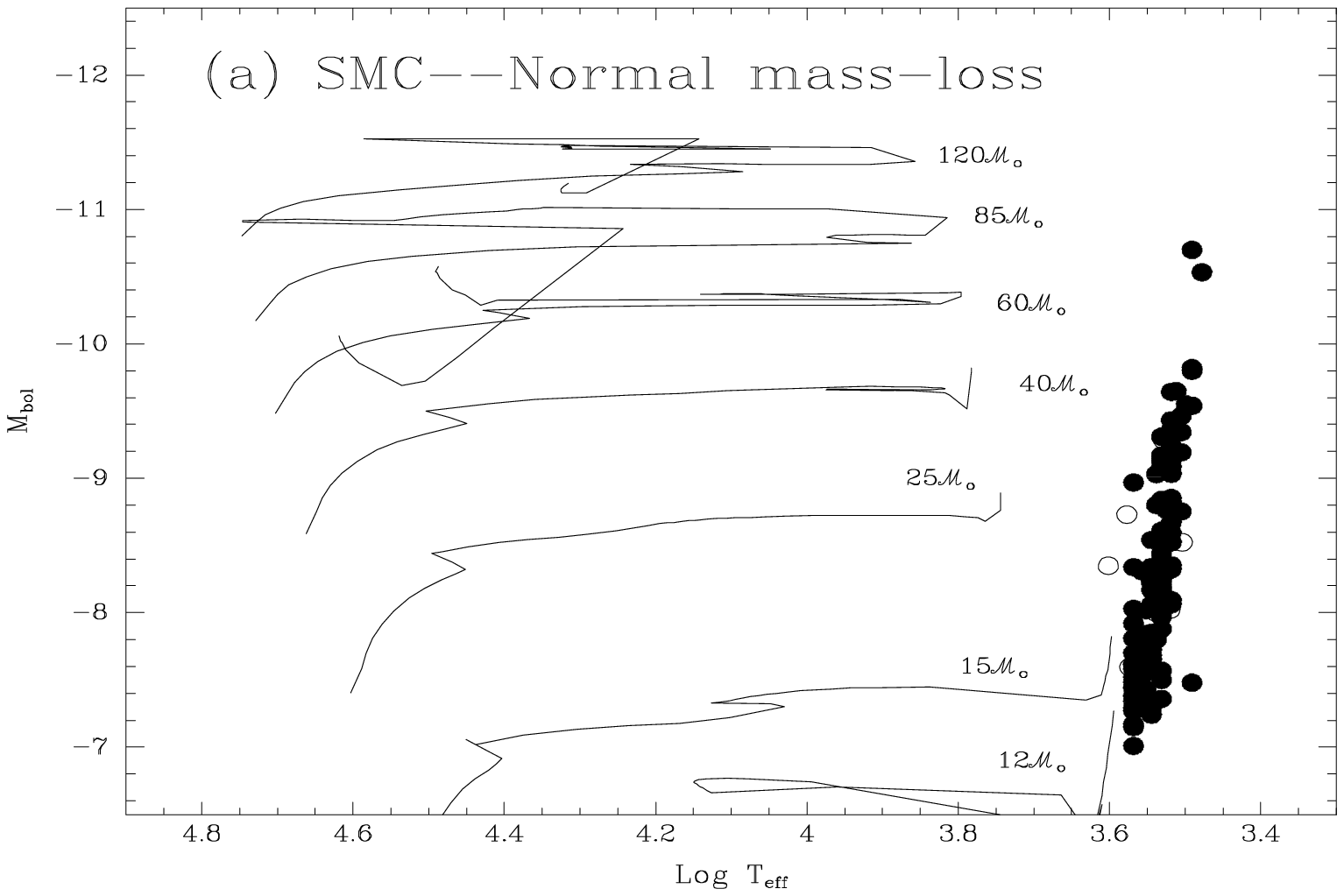,width=5.0truein}
\psfig{figure=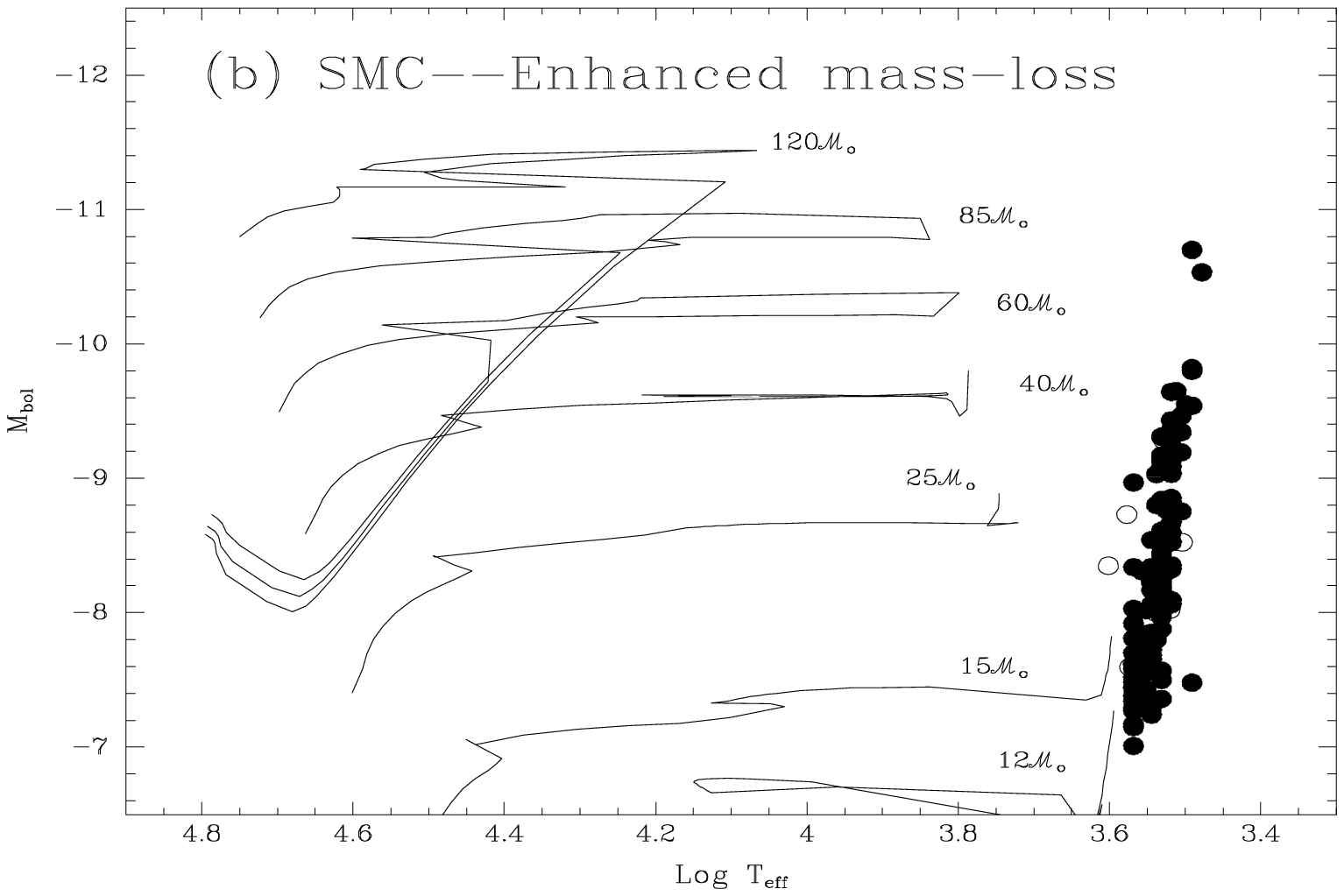,width=5.0truein}
\caption{The evolutionary tracks do not actually go sufficiently far to the right to produce RSGs as cool and as luminous as what is observed.  Note too that the tight relation of RSGs in these diagrams, with stars of
lower luminosities being of slightly earlier (hotter) spectral types.  The figure is based upon data from Massey \& Olsen (2003). \label{fig:rsgs}
}
\end{figure}

This work has revealed a potentially serious drawback to stellar evolutionary
models, namely that none of the ``modern" stellar evolutionary models produce
RSGs as cool and luminous as what is actually observed.  This is illustrated
in Figure~\ref{fig:rsgs}.  We see too that there is a tight relationship
between luminosity and effective temperature of the RSGs.  Matching this
will provide a challenge to future stellar evolutionary models.

\begin{figure}
\psfig{figure=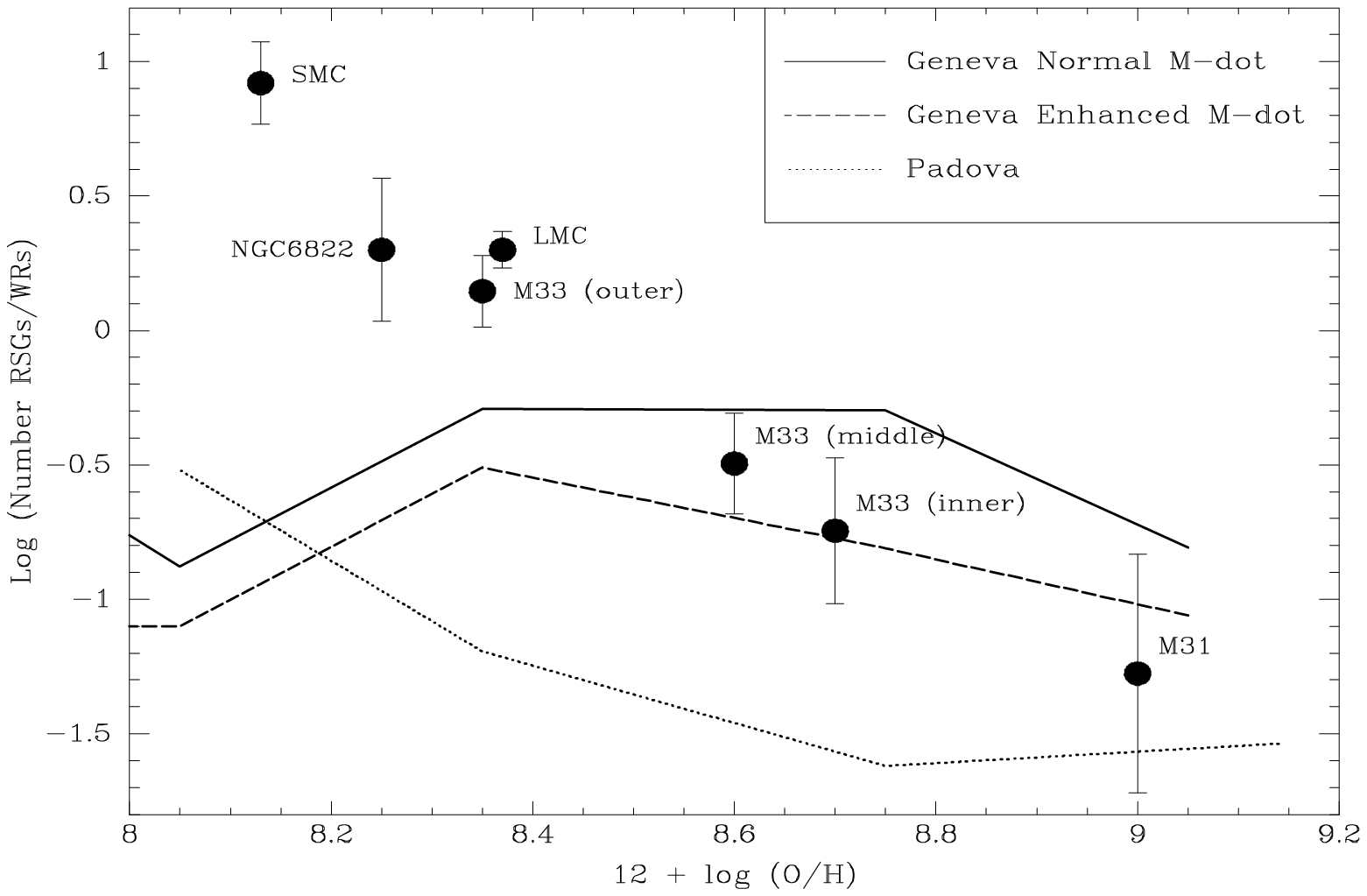,width=5.0truein}
\caption{Although the relative number of RSGs and WRs show a very strong
trend with metallicity, none of the current (non-rotation) evolutionary
models reproduce this trend. \label{fig:wrsrsgs}
}
\end{figure}

Finally we note that the relative number of RSGs and WRs show a very strong
trend with metallicity, changing by two orders of magnitude over just 
0.8~dex in metallicity (Figure~\ref{fig:wrsrsgs}.  Again, none of the
stellar evolutionary models reproduce this trend.

\subsection{Coeval Tests}
\label{Sec:coeval}

So far the tests described have involved mixed-aged populations, where we
compare the relative number of WCs to WNs, or RSGs to WRs.  But, in 
principle a far more sensitive test is possible, which involves using
coeval regions to determine the progenitor masses as a function of
metallicity.

This is a classical method first used by Sandage (1953) to determine the masses
of RR Lyrae stars using globular clusters, and subsequently used by
Anthony-Twarog (1982) to find the progenitor masses
of white dwarfs in intermediate-age clusters.  However, it is one thing do
do this for clusters that are $10^{10}$ yr old, or even
40-70~Myr old,
and quite another to do it for clusters that are only a few million years
old: the degree of coevality required is quite high.  However, recall that
Hillenbrand et al.~(1993) study of NGC 6611 
(discussed in 
Section~\ref{Sec:SF}
)
found that most of the massive stars were born over a very short period of
time ($<1$~Myr).  The data used to construct the H-R diagrams themselves
can be used to answer the degree of coevality.

\begin{figure}
\psfig{figure=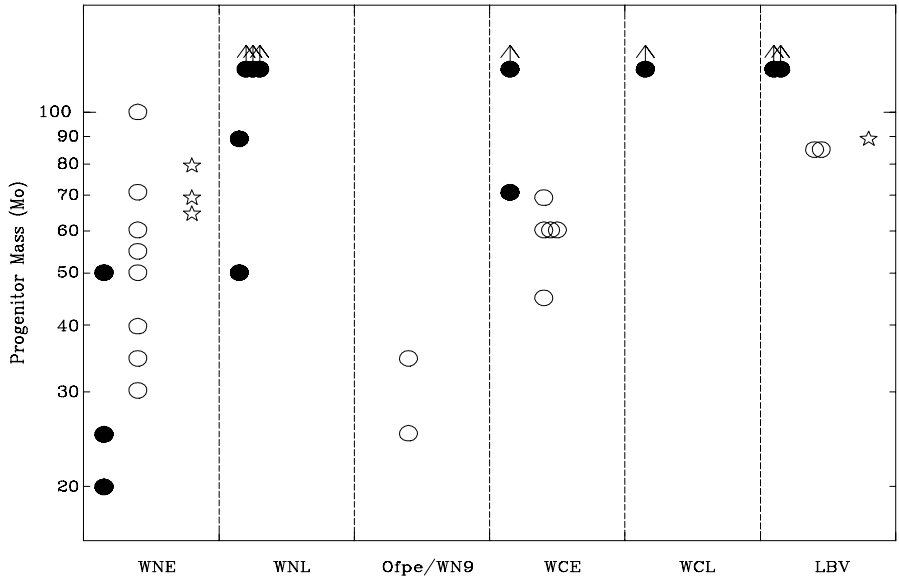,width=5.0truein}
\caption{ The progenitor masses are shown for various types of evolved massive stars for the SMC (stars), LMC (open circles), and Milky Way (filled circles).  We see that LBVs come from only the most massive stars.
The Ofpe/WN9s come from lower mass stars.  Of the WNEs, the progenitor masses are highest in the SMC, span a larger range to lower masses in the LMC, and are even lower in the Milky Way, as one might expect.  The figure is
is modified from one that appears in Massey et al.\ (2001).\label{fig:kathy}
}
\end{figure}

Massey, Waterhouse, \& DeGioia-Eastwood (2000) and Massey, DeGioia-Eastwood,
\& Waterhouse (2001) selected 19 clusters in the Magellanic Clouds and 12
clusters in the Milky Way that contained evolved massive stars, and obtained
photometry and spectroscopy of the most luminous unevolved cluster members.
About half of these met their stringent criteria for coevality, and hence
could be used to determine the progenitor masses of the evolved massive
stars using the masses of as-yet unevolved massive stars.  The results
are shown in Figure~\ref{fig:kathy}.

What we've learned from this is that LBVs all come from the highest mass
stars, as we expect.  The Ofpe/WN9s, which have sometimes been linked to
the LBVs, do not come from the highest mass stars---instead, their progenitor
masses are quite low.  Interestingly, are strong trend in the progenitor
masses of WN stars in the three galaxies: in the SMC only the highest mass
stars become WRs, apparently, which is consistent with the SMC's low
metallicity and the expectations of the Conti scenario that the amount of
mass-loss will determine the evolution of massive stars.  In the LMC, a larger
range of progenitor masses yield WNs.  And, in the Milky Way, stars from
masses as low as 20$\cal M_\odot$, up to that of $>>120\cal M_\odot$ become
WRs.

\section{What's Next?}
\label{Sec:Next}

Studies of the massive star content of nearby galaxies (including our own!)
have shed important light on star-formation and the evolution of massive
stars.  What comes next?

\begin{itemize}
\item We need to move beyond the Magellanic Cloud st complete galaxy-wide
surveys for WRs, RSGs, and BSGs in M31, M33, NGC 6822, IC10, and other
galaxies of the Local Group.

\item Follow-up spectroscopy with 8-m class telescopes will allow meaningful
H-R diagrams to be constructed, allowing careful tests of models as a
function of metallicity.

\item Coeval studies need to be extended to higher metallicity systems, such as 
M31

\end{itemize}

\begin{acknowledgments}
This contribution has been supported by the National Science Foundation,
through grant AST0093060.

\end{acknowledgments}

\vfill\noindent
{\it  The Local Group as an Astrophysical Laboratory,} 2003 STScI May
Symposium, eds. M. Livio et al., (Cambridge:  Cambridge U. Press)

\end{document}